\pdfoutput=1

\documentclass[12pt,a4paper,hyperref]{JHEP3}

\let\ifpdf\relax
\usepackage{ifpdf}
\usepackage{color}
\let\normalcolor\relax
\usepackage{mathtools}
\usepackage{shuffle}
\usepackage{cite}

\renewcommand{\tilde}{\widetilde}

\newcommand{\eq}[1]{\vspace{-0.175cm}\begin{equation}#1\vspace{-0.175cm}\end{equation}}

\newcommand{\mi}{\,{\rm\rule[2.4pt]{6pt}{0.65pt}}\,}
\newcommand{\pl}{\hspace{0.5pt}\text{{\small+}}\hspace{0.75pt}}
\newcommand{\ab}[1]{\langle #1 \rangle}

\definecolor{perm}{rgb}{0.1,0.45,0.85}
\definecolor{deemph}{rgb}{0.5,0.5,0.5}
\definecolor{bridge8}{rgb}{0.2,0,1.00}
\definecolor{bridge7}{rgb}{0.184615,0,0.923}
\definecolor{bridge6}{rgb}{0.184615,0,0.923}
\definecolor{bridge5}{rgb}{0.153846,0,0.769}
\definecolor{bridge4}{rgb}{0.138462,0,0.692}
\definecolor{bridge3}{rgb}{0.123077,0,0.615}
\definecolor{bridge2}{rgb}{0.107692,0,0.538}
\definecolor{bridge1}{rgb}{0.0923077,0,0.462}
\definecolor{bridge0}{rgb}{0,0,0}
\definecolor{iden}{rgb}{0.75,0.2,0}
\definecolor{hred}{rgb}{0.8,0.1,0.3}
\definecolor{hblue}{rgb}{0.2,0,0.8}
\definecolor{hgreen}{rgb}{0.0,0.4,0.2}
\definecolor{hteal}{rgb}{0.0,0.545,0.7451}
\newcommand{\fwbox}[2]{\text{\makebox[#1][c]{$\hspace{-150pt}\displaystyle#2\hspace{-150pt}$}}}
\newcommand{\fwboxL}[2]{\text{\makebox[#1][l]{$#2$}}}

\definecolor{hred2}{rgb}{0.82,0.216,0.435}

\title{\hspace{-0.0cm}\mbox{{\LARGE On-Shell Structures of MHV Amplitudes}}\\{\LARGE Beyond the Planar Limit}}
\author{\vspace{-.5cm}Nima Arkani-Hamed$^{a}$, Jacob L. Bourjaily$^{b}$, Freddy Cachazo$^{c}$, Alexander Postnikov$^{d}$, and Jaroslav Trnka$^{e}$\\
{\footnotesize
\mbox{\hspace{-15pt}{\it $^{\!\!\phantom{f}a}$School of Natural Sciences, Institute for Advanced Study, Princeton, NJ}}\\
\mbox{\hspace{-15pt}{\it $^{\!\!\phantom{f}b}$Niels Bohr International Academy and Discovery Center, Copenhagen, Denmark}}\\
\mbox{\hspace{-15pt}{\it $^{\!\!\phantom{f}c}$Perimeter Institute for Theoretical Physics, Waterloo, Ontario, Canada}}\\
\mbox{\hspace{-15pt}{\it $^{\!\!\phantom{f}d}$Department of Mathematics, Massachusetts Institute of Technology, Cambridge, MA}}\\
\mbox{\hspace{-15pt}{\it $^{\!\!\phantom{f}e}$Walter Burke Institute for Theoretical Physics, California Institute of Technology, Pasadena, CA}}}\vspace{-.5cm}}
\abstract{We initiate an exploration of on-shell functions in $\mathcal{N}\!=\!4$ SYM beyond the planar limit by  providing compact, combinatorial expressions for all leading singularities of MHV amplitudes and showing that they can always be expressed as a positive sum of differently ordered Parke-Taylor tree amplitudes. This is understood in terms of an extended notion of positivity in $G(2,n)$, the Grassmannian of $2$-planes in \mbox{$n$ dimensions}: a single on-shell diagram can be associated with many different ``positive" regions, of which the familiar $G_+(2,n)$ associated with planar diagrams is just one example. The decomposition into Parke-Taylor factors is simply a ``triangulation" of these extended positive regions. The $U(1)$ decoupling and Kleiss-Kuijf (KK) relations satisfied by the Parke-Taylor amplitudes also follow naturally from this geometric picture. These results suggest that non-planar MHV amplitudes in $\mathcal{N}\!=\!4$ SYM at all loop orders can be expressed as a sum of polylogarithms weighted by color factors and (unordered) Parke-Taylor amplitudes.}
\preprint{CALT-TH-2014-169}

\begin{document}
\section{Introduction}\label{introduction_section}

While recent years have seen tremendous advances in our understanding of scattering amplitudes for $\mathcal{N}\!=\!4$ SYM in the planar limit, much less is known beyond the planar limit (despite some impressive inroads for four- and five-particle amplitudes---see e.g.\ \cite{Bern:2010tq,Carrasco:2011mn}). And yet there are many indications that there is an even richer structure waiting to be unearthed beyond the planar limit. For instance the amplitudes for different color-orderings satisfy non-trivial relations between each other---such as the $U(1)$ decoupling and Kleiss-Kuijf (KK) identities, and the BCJ relations \cite{Bern:2008qj,Bern:2010ue,Bern:2010fy}---which cannot be understood within each color ordering separately; and similarly, the remarkable relations between Yang-Mills and gravitational amplitudes can not be seen in the planar limit.

The Grassmannian representation of planar $\mathcal{N}\!=\!4$ scattering amplitudes \cite{ArkaniHamed:2009dn,ArkaniHamed:2012nw} is based on a striking physical principle: all amplitudes in the theory can be directly represented as on-shell processes, where the fundamental $3$-particle amplitudes are glued together with all intermediate particles taken to be on-shell, eliminating any reference to ``virtual particles". Of course, on-shell diagrams have long been known to have direct physical interpretation as ``cuts" which compute discontinuities of loop amplitudes across branch-cuts. But this connections exists much more generally: on-shell diagrams in any theory with massless particles are most naturally computed and understood in terms of an associated structure in the Grassmannian defined by its $3$-particle amplitudes. The additional novelty described in \cite{ArkaniHamed:2010kv,ArkaniHamed:2012nw} is that planar loop-amplitude integrands can be directly represented by on-shell diagrams. While we do not yet know whether this is possible in general, the successes of this picture in planar $\mathcal{N}\!=\!4$ SYM motivates us to try and find a completely on-shell description of the physics for general theories.

The main obstacle to finding an on-shell representation of $\mathcal{N}\!=\!4$ SYM beyond the planar limit is a simple, almost kinematical one: that the notion of ``the" integrand for scattering amplitudes becomes ambiguous beyond the planar limit. But we can certainly begin to explore on-shell diagrams beyond the planar limit, which minimally have the physical interpretation of leading singularities. Any on-shell diagram is associated with some region in the Grassmannian together with a volume form that has logarithmic singularities along its boundary. Recent studies \cite{Arkani-Hamed:2014via,ThreeLoop} have found evidence that loop integrands can be represented in a ``$d\!\log$"-form, with manifestly logarithmic singularities in loop-momentum space. These two facts support the hope that some representation of an integrand can be given in purely on-shell terms, making the logarithmic singularities of amplitudes manifest at the integrand-level.

All of these considerations strongly motivate an investigation of non-planar on-shell diagrams in $\mathcal{N}\!=\!4$ SYM. We will consider an arbitrary diagram obtained by gluing black and white 3-particle vertices together. Any diagram is associated with an obvious color factor. The color-Jacobi relations will undoubtedly be important for the actual amplitudes, and for the hope of an on-shell representation of them. However, they factor out of each on-shell diagram, and we will omit them in what follows. In this note, we will further specialize to on-shell diagrams for MHV amplitudes, which are especially simple but already reveal essential new features that occur for non-planar amplitudes more generally.

In \mbox{section \ref{classification_subsection}} we show that all leading singularities can be characterized by a list of $(n\mi2)$ subsets of external leg labels; and in \mbox{section \ref{grassmannian_formula_subsection}} we show how this data directly encodes the corresponding function. Analyzing the singularities of these functions geometrically in \mbox{section \ref{generalized_positivity_section}} will lead us to an extended notion of positivity in the Grassmannian and allow us to discover the remarkable fact that all leading singularities of MHV amplitudes can be expressed as the positive sum (with all coefficients $+1$) of {\it planar} leading singularities. These considerations provide a geometric basis for the $U(1)$ decoupling identities and the Kleiss-Kuijf (KK) relations \cite{Kleiss:1988ne} which relate amplitudes involving differently ordered sets of external legs.

If all non-planar loop amplitude integrands in $\mathcal{N}\!=\!4$ SYM are in fact logarithmic as conjectured in \mbox{ref.\ \cite{Arkani-Hamed:2014via}}, this suggests that all MHV loop amplitudes of $\mathcal{N}\!=\!4$ SYM can be expressed in terms of polylogarithms with all coefficients being color factors times planar tree amplitudes involving differently ordered sets of external legs.

\newpage
\section{\mbox{General Leading Singularities of MHV Amplitudes}}\label{mhv_diagrams_section}\vspace{-0pt}
\subsection{General Leading Singularities and the Reduction of Diagrams}

Leading singularities correspond to on-shell diagrams obtained by taking successive residues of loop amplitude integrands---putting some number of internal particles on-shell. The corresponding on-shell functions encode the coefficients of transcendental functions appearing in the loop expansion and thereby capture much about the structure of scattering amplitudes beyond the leading order of perturbation theory.

Because all leading singularities are on-shell diagrams, they can be computed in terms of an auxiliary Grassmannian integral as described in \mbox{ref.\ \cite{ArkaniHamed:2012nw}}. For $\mathcal{N}\!=\!4$ SYM, the volume-form on the auxiliary Grassmannian is just the product of $d\!\log$'s of edge variables. Naively, the leading singularities could be as complicated as loop diagrams, with an infinite number of objects at higher and higher loops. At high enough loop-order, however, the cut propagators don't localize all the internal momenta, and we have a non-trivial form in the remaining variables. We can take residues on enough of these to fully localize them to the leading singularities.
Naively, there are an infinite number of objects of this form, descending from taking residues on arbitrarily complicated on-shell diagrams. However, the Grassmannian representation makes an otherwise highly non-trivial fact completely obvious: the number of leading singularities for any $(n,k)$ is finite.

Starting from an arbitrary on-shell diagram, all but a small number of edges can always be removed (via residues setting edge weights to zero) without encountering any enhanced constraints on the external kinematical data. This is known as reduction, and the finiteness of the set of reduced diagrams follows from simple dimensional arguments. We may define a graph to be reduced if the dimension of its boundary measurement matrix (as a submanifold of $G(k,n)$) is equal to the number of independent edge weights. Unlike for the planar theory, it is not always possible to transform a diagram using square moves and mergers to expose a bubble somewhere in the diagram; this is related to another important difference with the planar case: a non-reduced diagram can sometimes be reduced in inequivalent ways. Both of these novelties are illustrated in the following example:
\vspace{-5pt}\eq{\raisebox{-76pt}{\includegraphics[scale=1]{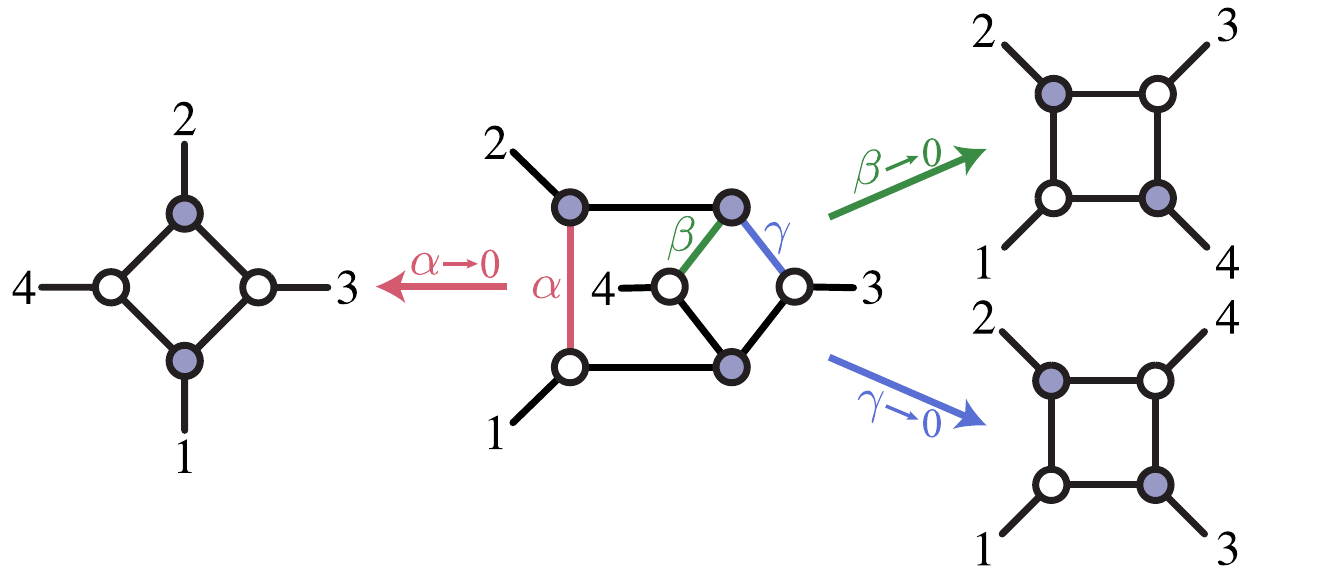}}\vspace{-5pt}}

For the rest of this work, we will focus our attention on the case of reduced on-shell diagrams relevant to MHV amplitudes. We will describe in \mbox{section \ref{classification_subsection}} how all such diagrams can be classified combinatorially, and show how this combinatorial data can be used to directly construct the corresponding function in \mbox{section \ref{grassmannian_formula_subsection}}.

\subsection{Classification of Reduced, MHV On-Shell Functions and Diagrams}\label{classification_subsection}
We begin by showing that any generically non-vanishing, reduced MHV on-shell diagram is naturally labeled by a list of $(n\mi2)$ triples of external legs---corresponding to the labels of the (precisely) three external legs attached (via white vertices) to each of the $(n\mi2)$ black vertices in the diagram. This labeling is illustrated in the following example involving 9 particles:
\eq{\raisebox{-65pt}{\includegraphics[scale=0.9]{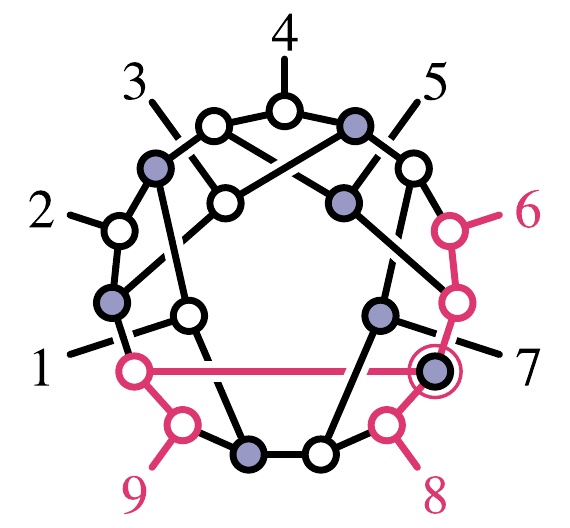}}\raisebox{-3pt}{{\huge$\Leftrightarrow$}}\left\{\!\begin{array}{@{$($}c@{$\,$}c@{$\,$}c@{$)$}}1&2&4\\1&8&9\\2&9&3\\3&6&4\\4&6&5\\6&8&7\\\multicolumn{1}{@{${\color{hred2}(}$}c@{$\,$}}{{\color{hred2}6}}&{\color{hred2}8}&\multicolumn{1}{@{}c@{${\color{hred2})}$}}{{\color{hred2}9}}\end{array}\!\right\}}
Notice that each of the $7\!=\!(n\mi2)$ black vertices is labeled by three external legs according to the rule described above.

The proof that all non-vanishing leading singularities of MHV amplitudes can be labeled in this way is fairly straightforward. We can characterize an arbitrary on-shell diagram involving $n_V\!\equiv\!n_B\pl n_W$ vertices and $n_I$ internal edges by defining:
\eq{k\equiv 2n_B+n_W-n_I,\quad\mathrm{and}\quad n_\delta\equiv 4n_V-3n_I-4,\label{definitions_of_k_and_n_delta}}
where diagrams related to MHV amplitudes have $k\!=\!2$, and $n_\delta$ counts the number of \mbox{$\delta$-function} constraints imposed on the external kinematics beyond overall momentum conservation; for a reduced diagram---an ordinary function of the kinematical data---we have $n_\delta\!=\!0$. The number of external legs of any trivalent graph is given by,
\eq{n\equiv3n_V-2n_I,}
from which we see that $n_\delta\!=\!n\pl n_V\mi n_I\mi 4$. Using this, we see that any on-shell diagram with $k\!=\!2\!=\!n_B\pl n_V\mi n_I$ and $n_\delta\!=\!0$ will involve $n_B\!=\!n\mi2$ black vertices.

Without loss of generality, we may suppose that every external leg attaches to the graph at a white vertex---by adding bivalent white vertices to each external leg if necessary. Observe that removing a white-to-white edge lowers both $n_W$ and $n_I$ by one; therefore, $k\!=\!2n_B\pl n_W\mi n_I$ is left invariant by collapsing all internal trees of white vertices. Let us suppose that this has been done, and let $n_{W'}$ denote the number of white multi-vertices. For on-shell functions with kinematical support, no two legs can be connected to the same white vertex; this requires that $n\!\geq n_{W'}$.

Let us now show that $n\!=\!n_{W'}$, which implies that there are no black-to-black internal edges. Let us say the number of white multi-vertices is $n\pl q$; we want to show that $q\!=\!0$. From the definition of $k$,
\eq{k=2n_B+n_{W'}-n_I=2n_B+(n+q)-n_I=3n_B+2+q-n_I,}
from which we see that for $k\!=\!2$, $3n_B\!=\!n_I\mi q$. But $3n_B\!\geq\!n_I$ on general grounds, and so we must have that $q\!=\!0$, and hence $3n_B\!=\!n_I$. Because $q\!=\!0$, there is one leg connected to each white multi-vertex ($n_{W'}\!=\!n$); and because $3n_B\!=\!n_I$, there can be no black-to-black internal edges. Thus every black vertex connects to precisely three external legs via white multi-vertices, as we wanted to prove.

Therefore, any MHV ($k\!=\!2$) on-shell diagram corresponding to an ordinary function of the external data $(n_\delta\!=\!0)$ with kinematical support will involve precisely $(n\mi2)$ black vertices, each of which is attached to exactly three external legs via white vertices. Thus, we can label any such diagram by a set $T$ consisting of triplets $\tau\!\in\!T$ of leg labels for each of the $(n\mi2)$ black vertices.

We can illustrate how this labeling works with the following examples:
\eq{\begin{split}\hspace{-210pt}\fwbox{135pt}{\hspace{-20pt}\raisebox{-33pt}{\includegraphics[scale=.9]{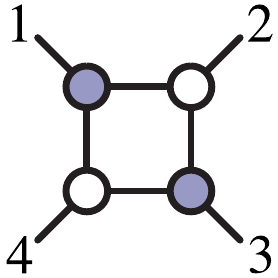}}}\left\{\!\begin{array}{@{$($}c@{$\,$}c@{$\,$}c@{$)$}}1&2&4\\2&3&4\end{array}\!\right\}\hspace{20pt}\fwbox{135pt}{\raisebox{-42.75pt}{\includegraphics[scale=.9]{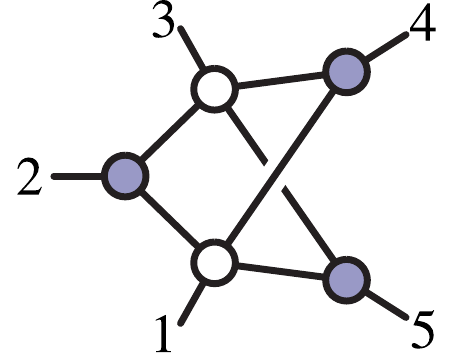}}}\left\{\!\begin{array}{@{$($}c@{$\,$}c@{$\,$}c@{$)$}}1&2&3\\1&3&4\\1&3&5\end{array}\!\right\}\hspace{-200pt}\end{split}\label{example_on_shell_diagrams_1}}
\eq{\begin{split}\hspace{-210pt}\fwbox{135pt}{\raisebox{-65pt}{\includegraphics[scale=0.9]{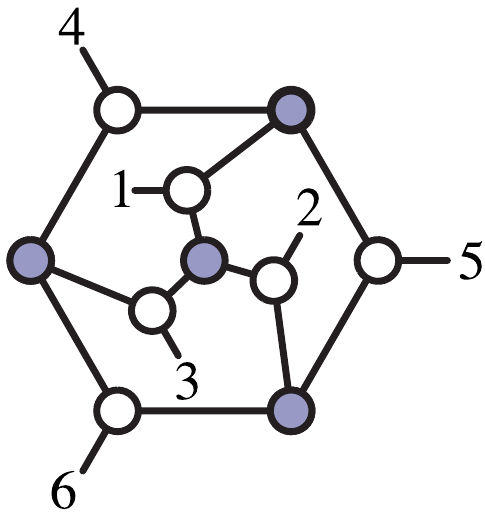}}}\left\{\!\begin{array}{@{$($}c@{$\,$}c@{$\,$}c@{$)$}}1&2&3\\2&5&6\\3&4&6\\4&5&1\end{array}\!\right\}\hspace{20pt}\fwbox{135pt}{\raisebox{-65pt}{\includegraphics[scale=0.9]{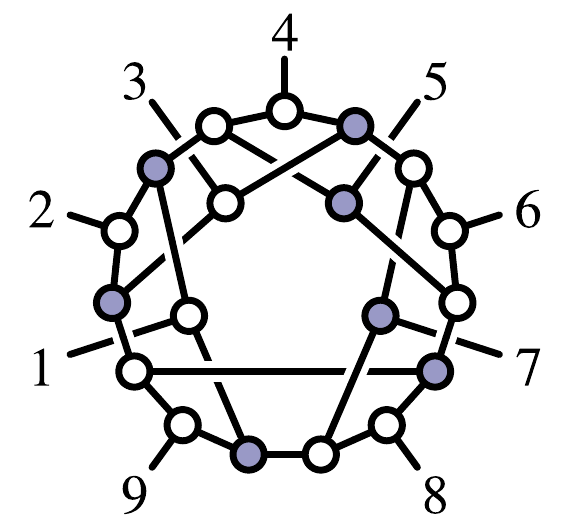}}}\!\!\left\{\!\begin{array}{@{$($}c@{$\,$}c@{$\,$}c@{$)$}}1&2&4\\1&8&9\\2&9&3\\3&6&4\\4&6&5\\6&8&7\\6&9&8\end{array}\!\right\}\hspace{-200pt}\end{split}\label{example_on_shell_diagrams_2}}

\noindent Notice that because there is no preferred way to order the external legs of a non-planar diagram, there is no preferred way to order the triples. And so while we have chosen a particular ordering for each triple in the examples above, these choices should be viewed as completely arbitrary.

\newpage
\subsection{Grassmannian Representations of MHV On-Shell Functions}\label{grassmannian_formula_subsection}
As described in \mbox{ref.\ \cite{ArkaniHamed:2012nw}}, any on-shell diagram $\Gamma$ of $\mathcal{N}\!=\!4$ SYM corresponds to an on-shell function $f_\Gamma$ that can be computed in terms of an auxiliary Grassmannian $C\!\in\!G(k,n)$ according to,
\vspace{0pt}\eq{f_\Gamma=\int\Omega_C\,\,\delta^{k\times4}\big(C\!\cdot\!\tilde{\eta}\big)\delta^{k\times2}\big(C\!\cdot\!\tilde\lambda\big)\delta^{2\times(n-k)}\big(\lambda\!\cdot\!C^{\perp}\!\big),\label{grassmannian_formula}}
where $\Omega_C$ is a volume form on the configuration $C$ with strictly logarithmic singularities. When expressed in terms of edge-variables $\alpha_i$ associated with the graph, $C(\alpha)$ is determined by boundary measurements, and \mbox{$\Omega_C\!=\!d\!\log(\alpha_1)\!\wedge\!\cdots\!\wedge d\!\log(\alpha_d)$}.

For any leading singularity of an MHV ($k\!=\!2$) amplitude, \mbox{$C(\alpha)$} is a $(2n\mi4)$-dimensional subspace, and so $\Omega_C$ is a top-dimensional form on $G(2,n)$. Thus, the $(2n\mi4)$ constraints \mbox{$\delta^{2\times(n-2)}\big(\lambda\!\cdot\!C^{\perp}\!\big)$} entirely localize the $2$-plane $C(\alpha)$ to be identical to the $2$-plane $\lambda$. Let us now describe how these constraints directly provide us with a formula for $f_\Gamma$ in terms of the triples $\tau\!\in\!T$ which label the graph.

Each black vertex of the diagram encodes a linear relation among the three external $\lambda$'s attached to it. Thus, a vertex labeled by the triple $(\tau_1\,\tau_2\,\tau_3)\!\in\!T$ is associated with the following contribution to (\ref{grassmannian_formula}):
\eq{(\tau_1\,\tau_2\,\tau_3)\raisebox{-1pt}{\text{{\Large$\;\Rightarrow\;$}}} \frac{1}{\mathrm{vol}(GL(1))}\frac{d\alpha_{\tau_1}}{\alpha_{\tau_1}}\frac{d\alpha_{\tau_2}}{\alpha_{\tau_2}}\frac{d\alpha_{\tau_3}}{\alpha_{\tau_3}}\delta^{2\times1}\big(\lambda_{\tau_1}\alpha_{\tau_1}+\lambda_{\tau_2}\alpha_{\tau_2}+\lambda_{\tau_3}\alpha_{\tau_3}\big),\label{black_vertex_contribution}}
where ``$1/\mathrm{vol}(GL(1))$'' is an instruction to eliminate any one of the redundant $\alpha$'s. (The reason for introducing such a redundancy here is that it makes all the singularities arising from the vertex manifest.)

These $\delta$-functions are very easy to integrate, as each encodes an instance of the unique three-term identity satisfied by generic two-dimensional vectors:\\[-8pt]
\eq{\lambda_{\tau_1}\ab{\tau_2\,\tau_3}+\lambda_{\tau_2}\ab{\tau_3\,\tau_1}+\lambda_{\tau_3}\ab{\tau_1\,\tau_2}=0;}
from this, we see that on the support of the $\delta$-functions,
\eq{\{\alpha_{\tau_1},\alpha_{\tau_2},\alpha_{\tau_3}\}\mapsto\{\alpha_{\tau_1}^*,\alpha_{\tau_2}^*,\alpha_{\tau_3}^*\}=\{\ab{\tau_2\,\tau_3},\ab{\tau_3\,\tau_1},\ab{\tau_1\,\tau_2}\}.}
Because this argument is repeated for each triple $\tau\!\in\!T$, it provides us with an explicit representative of the plane $C^\perp(\vec{\alpha}^*)$ satisfying the constraints $\delta^{2(n-2)}\big(\lambda\!\cdot\!C^{\perp}(\vec{\alpha})\big)$.

Combining the contributions to (\ref{grassmannian_formula}) from each black vertex, we find the following expression for the leading singularity:
\eq{f_\Gamma=\prod_{\tau\in T}\left(\frac{1}{\ab{\tau_1\tau_2}\ab{\tau_2\tau_3}\ab{\tau_3\,\tau_1}}\right)\delta^{2\times4}\big(C(\vec{\alpha}^*)\!\cdot\!\widetilde\eta\big)\delta^{2\times2}\big(C(\vec{\alpha}^*)\!\cdot\!\widetilde\lambda\big).\label{pre_gauge_fixed_form}}
Notice, however, that the $\delta$-functions in this expression imposing supermomentum conservation are not written in their standard form. Indeed, although $C^{}(\vec{\alpha}^*)$ and $\lambda$ are equivalent as $2$-planes in $n$ dimensions, they can differ by a $GL(2)$ transformation. (Once we have chosen a Lorentz-frame, $\lambda$ does not correspond to a `$2$-plane', but rather a $(2\!\times\!n)$ matrix.) To bring the expression (\ref{pre_gauge_fixed_form}) to its standard form requires a ``gauge-fixing'' factor---a Jacobian---$\Psi$ relating the two:
\eq{\delta^{2\times4}\big(C(\vec{\alpha}^*)\!\cdot\!\widetilde\eta\big)\delta^{2\times2}\big(C(\vec{\alpha}^*)\!\cdot\!\widetilde\lambda\big)\equiv\Psi\,\,\delta^{2\times4}\big(\lambda\!\cdot\!\widetilde\eta\big)\delta^{2\times2}\big(\lambda\!\cdot\!\widetilde\lambda\big).\label{formula_up_to_gauge_fixing}}

To determine the gauge-fixing factor $\Psi$, we need to be more explicit about how $C^\perp(\vec{\alpha}^*)$ is related to the matrix $\lambda^\perp$. An explicit representative of $\lambda^\perp$ can be obtained by choosing to expand all the $\lambda$'s as $2$-vectors in terms of a basis consisting of some pair of external $\lambda$'s, $\{\lambda_a,\lambda_b\}$; this choice of coordinates for $\lambda$ induces a Jacobian $\ab{a\,b}^{-2}$ which contributes to $\Psi$. Having chosen this gauge for $\lambda$, $\lambda^\perp$ is completely fixed to have its $(n\mi2)$ columns in the complement of $\{a,b\}$ equal to the identity matrix. But importantly, the representative matrix $C^\perp(\vec{\alpha}^*)$ constructed above is not in this gauge-fixed form; and so we must multiply it on the left by the inverse of the matrix, denoted $M_{ab}$, obtained from $C^{\perp}(\vec{\alpha}^*)$ by deleting its columns $\{a,b\}$. This gauge-fixing induces a Jacobian of $\det(M_{ab})^2$ which constitutes the final contribution to $\Psi$. Thus, having made the (arbitrary) choice of $\{\lambda_a,\lambda_b\}$ as a basis for the $\lambda$'s, the complete gauge-fixing factor $\Psi$ of (\ref{formula_up_to_gauge_fixing}) is found to be:\\[-8pt]
\eq{\Psi=\left(\det\big(M_{ab}\big)/\ab{a\,b}\right)^2.\label{gauge_fixing_factor}}
(It can be shown that $\Psi$ is independent of the choice of $\{a,b\}$.)

Putting everything together, we find the following expression for any on-shell diagram $\Gamma$ labeled by triples $\tau\!\in\!T$:
\eq{\framebox[0.6\textwidth]{$\displaystyle f_\Gamma=\frac{\left(\det\big(M_{ab}\big)/\ab{a\,b}\right)^2}{\prod_{\tau\in T}\ab{\tau^{}_1\tau^{}_2}\ab{\tau^{}_2\tau^{}_3}\ab{\tau^{}_3\tau^{}_1}}\delta^{2\times4}\big(\lambda\!\cdot\!\widetilde\eta\big)\delta^{2\times2}\big(\lambda\!\cdot\!\widetilde{\lambda}\big).$}\label{freddys_formula_for_general_diagrams}}
This formula should really be viewed as defining a differential form on the space of external kinematical data, $\widetilde{f}_\Gamma$, which is constrained by supermomentum conservation:\\[-8pt]
\eq{f_\Gamma\equiv\widetilde{f}_\Gamma\times\delta^{2\times4}\big(\lambda\!\cdot\!\widetilde\eta\big)\delta^{2\times2}\big(\lambda\!\cdot\!\widetilde{\lambda}\big).}
To simplify our expressions below, we will leave supermomentum conservation implicit and write $\widetilde{f}_\Gamma$ in the particular examples discussed.

Let us see how this formula looks for some of the examples given in (\ref{example_on_shell_diagrams_1}) and (\ref{example_on_shell_diagrams_2}). For the leading singularity involving $5$ particles in (\ref{example_on_shell_diagrams_1}) we have,
\vspace{-0pt}\eq{\scalebox{1}{$\hspace{-202pt}\fwboxL{145pt}{\raisebox{-42.75pt}{\includegraphics[scale=0.9]{g25_graph}}}\!\!\!\left\{\!\begin{array}{@{$($}c@{$\,$}c@{$\,$}c@{$)$}}1&2&3\\1&3&4\\1&3&5\end{array}\!\right\}\!\!\raisebox{-1.pt}{\text{{\Large$\;\Rightarrow\;$}}}\fwboxL{225pt}{C^{\perp}(\vec{\alpha}^*)\equiv\!\begin{array}{@{}c@{}}\begin{array}{@{}c@{$\,$}c@{$\,$}c@{$\,$}c@{$\,$}c@{}}\fwbox{22.5pt}{1}&\fwbox{22.5pt}{2}&\fwbox{22.5pt}{3}&\fwbox{22.5pt}{4}&\fwbox{22.5pt}{5}\\\hline\end{array}\\\left(\begin{array}{@{}c@{$\,$}c@{$\,$}c@{$\,$}c@{$\,$}c@{}}\fwbox{22.5pt}{\ab{2\,3}}&\fwbox{22.5pt}{\ab{3\,1}}&\fwbox{22.5pt}{\ab{1\,2}}&\fwbox{22.5pt}{0}&\fwbox{22.5pt}{0}\\\ab{3\,4}&0&\ab{4\,1}&\ab{1\,3}&0\\\ab{3\,5}&0&\ab{5\,1}&0&\ab{1\,3}\end{array}\right),\\~\end{array}}\hspace{-210pt}$}\nonumber\vspace{-0pt}}
from which we see that $\Psi\!=\!\ab{3\,1}^4$, yielding the expression:
\vspace{-0pt}\eq{\hspace{-278pt}\raisebox{-42.75pt}{\includegraphics[scale=.9]{g25_graph}}\hspace{-25pt}\begin{array}{@{}l@{}l@{}}\hspace{20pt}=\displaystyle\frac{\ab{3\,1}\fwboxL{0pt}{}}{\ab{1\,2}\ab{2\,3}\ab{3\,4}\ab{4\,1}\ab{3\,5}\ab{5\,1}}\,.\phantom{\delta^{2\times4}\big(\lambda\!\cdot\!\widetilde{\eta}\big)\delta^{2\times2}\big(\lambda\!\cdot\!\widetilde{\lambda}\big)}\end{array}\hspace{-220pt}\label{g25_example_formula}\vspace{-0pt}}
\noindent Following the same procedure for the $6$-particle example of (\ref{example_on_shell_diagrams_2}), we see that
\vspace{-0pt}\eq{\scalebox{1}{$\hspace{-220pt}\fwbox{145pt}{\raisebox{-61pt}{\includegraphics[scale=0.85]{g26_graph}}}\!\!\!\left\{\!\begin{array}{@{$($}c@{$\,$}c@{$\,$}c@{$)$}}1&2&3\\2&5&6\\3&4&6\\4&5&1\end{array}\!\right\}\!\!\raisebox{-1.pt}{\text{{\Large$\;\Rightarrow\;$}}}\fwboxL{225pt}{C^{\perp}(\vec{\alpha}^*)\equiv\!\begin{array}{@{}c@{}}\begin{array}{@{}c@{$\,$}c@{$\,$}c@{$\,$}c@{$\,$}c@{$\,$}c@{}}\fwbox{22.5pt}{1}&\fwbox{22.5pt}{2}&\fwbox{22.5pt}{3}&\fwbox{22.5pt}{4}&\fwbox{22.5pt}{5}&\fwbox{22.5pt}{6}\\\hline\end{array}\\\left(\begin{array}{@{}c@{$\,$}c@{$\,$}c@{$\,$}c@{$\,$}c@{$\,$}c@{}}\fwbox{22.5pt}{\ab{2\,3}}&\fwbox{22.5pt}{\ab{3\,1}}&\fwbox{22.5pt}{\ab{1\,2}}&\fwbox{22.5pt}{0}&\fwbox{22.5pt}{0}&\fwbox{22.5pt}{0}\\0&\ab{5\,6}&0&0&\ab{6\,2}&\ab{2\,5}\\0&0&\ab{4\,6}&\ab{6\,3}&0&\ab{3\,4}\\\ab{4\,5}&0&0&\ab{5\,1}&\ab{1\,4}&0\end{array}\right),\\~\end{array}}\hspace{-210pt}\nonumber\vspace{-0pt}$}}
from which we see that $\Psi\!=\!\big(\!\ab{34}\ab{51}\ab{62}\pl\ab{14}\ab{25}\ab{63}\!\big)^2,$
allowing us to write:
\vspace{-0pt}{\small \eq{\hspace{-200pt}\raisebox{-61.5pt}{\includegraphics[scale=0.85]{g26_graph}}\raisebox{0pt}{$\displaystyle\hspace{5pt}=\frac{\big(\!\ab{34}\ab{51}\ab{62}\pl\ab{14}\ab{25}\ab{63}\!\big)^2\fwboxL{0pt}{\,\phantom{\delta^{2\times4}\big(\lambda\!\cdot\!\widetilde{\eta}\big)}}}{\ab{12}\ab{23}\ab{31}\ab{25}\ab{56}\ab{62}\ab{34}\ab{46}\ab{63}\ab{45}\ab{51}\ab{14}}.\phantom{\delta^{2\times2}\big(\lambda\!\cdot\!\widetilde{\lambda}\big)}$}\hspace{-200pt}\label{g26_example_formula}\vspace{-0pt}}
And finally, for the $9$-particle example of (\ref{example_on_shell_diagrams_2}), we find the formula:
\vspace{-5pt}{\small\eq{\hspace{-202pt}\!\!\!\raisebox{-30pt}{\includegraphics[scale=0.9]{g29_graph}}\hspace{-22pt}\raisebox{30pt}{$\displaystyle=\frac{\big(\!\ab{91}\hspace{-0pt}\ab{32}\hspace{-0pt}\ab{46}\mi\ab{16}\hspace{-0pt}\ab{43}\hspace{-0pt}\ab{29}\!\big)^2\fwboxL{00pt}{\,\phantom{\delta^{2\times4}\big(\lambda\!\cdot\!\widetilde{\eta}\big)\delta^{2\times2}\big(\lambda\!\cdot\!\widetilde\lambda\big)}}}{\ab{12}\ab{24}\ab{41}\ab{18}\ab{91}\ab{29}\ab{93}\ab{32}\ab{36}\ab{43}\ab{65}\ab{54}\ab{87}\ab{76}\ab{69}}.$}\label{g29_example}\hspace{-200pt}}}

\section{Extended Positivity in the Grassmannian}\label{generalized_positivity_section}

The formula given above for any MHV leading singularity, (\ref{freddys_formula_for_general_diagrams}), is remarkably concise and directly encodes all its singularities. But there is a beautiful fact that is obscured by such compact expressions: although the  collection of possible on-shell functions grows very rapidly in number and complexity with the number of external legs, it turns out that all such functions can be expanded as a positive sum---with all coefficients $1$---of {\it planar} on-shell functions involving differently ordered legs. This fact is best understood in terms of a generalized notion of positivity for the geometry of the auxiliary Grassmannian associated with the graph through equation (\ref{grassmannian_formula}).

As we have seen, the formula (\ref{freddys_formula_for_general_diagrams}) can be understood geometrically in terms of a top-dimensional volume form on the Grassmannian:
\eq{\Omega_C\equiv\prod_{\tau\in T}\frac{1}{\mathrm{vol}(GL(1))}\frac{d\alpha_{\tau_1}}{\alpha_{\tau_1}}\frac{d\alpha_{\tau_2}}{\alpha_{\tau_2}}\frac{d\alpha_{\tau_3}}{\alpha_{\tau_3}}=\prod_{\tau\in T}d\!\log\!\!\left(\!\frac{(\tau_1\,\tau_2)}{(\tau_3\,\tau_1)}\!\right)d\!\log\!\!\left(\!\frac{(\tau_2\,\tau_3)}{(\tau_3\,\tau_1)}\!\right),\label{dlog_top_form_formula}}
where $(a\,b)\!\equiv\!\det\{c_{a},c_{b}\}$. This differential form is entirely fixed by its logarithmic singularities, which can be thought to enclose some open region of the Grassmannian $G(2,n)$.

As described in \mbox{ref.\ \cite{ArkaniHamed:2012nw}}, given any {\it planar} on-shell diagram whose external legs are ordered in the clockwise direction around the graph, the configuration $C(\alpha)$ appearing in (\ref{grassmannian_formula}) is associated with the ``positive part" \cite{L1,Lusztig::1998,L2,P,KLS} of the (real) Grassmannian---where all the ordered minors $(a\,b)\!>\!0$ for $a\!<\!b$ when the edge variables $\alpha_i\!>\!0$. The form has logarithmic singularities on the boundaries of this positive region. Labeling the legs clockwise around the boundary of the graph by $\{1,2,\ldots,n\}$, it was shown in \mbox{ref.\ \cite{ArkaniHamed:2012nw}} that the canonical volume form $\Omega_C$ can always be written:
\eq{\Omega_C=\frac{d\alpha_1}{\alpha_1}\wedge\cdots\wedge\frac{d\alpha_{2n-4}}{\alpha_{2n-4}}=\frac{d^{2\times n}C}{\mathrm{vol}(GL(2))}\frac{1}{(1\,2)(2\,3)\cdots(n\,1)}.}
To see that this follows from positivity, observe that the codimension one boundaries of a generic positive configuration (viewed as an ordered set of points along $S^1$) correspond to configurations where two neighboring points collide, sending $\ab{a\,a\pl1}\!\to\!0$. Because the constraints $\delta^{2\times(n-2)}\big(\lambda\!\cdot\!C^{\perp}\big)$ in (\ref{grassmannian_formula}) localize $C\!\mapsto\!C^*\!\simeq\!\lambda$, we find the following expression for the on-shell function:
\vspace{8pt}\eq{\widetilde{f}_\Gamma=\frac{1}{\ab{1\,2}\ab{2\,3}\ab{3\,4}\cdots\ab{n\,1}}.\label{general_on_shell_form_planar}}

This on-shell function is the same for all reduced, planar on-shell diagrams with this ordering of the external legs. Moreover, (\ref{general_on_shell_form_planar}) is the complete ``tree-level'' MHV scattering amplitude for $n$ particles. This remarkably simple expression was first guessed by Parke and Taylor in \mbox{ref.\ \cite{Parke:1986gb}}, and so we will refer to the on-shell differential form (\ref{general_on_shell_form_planar}) as the `Parke-Taylor factor' for the specified ordering of external legs:
\eq{PT(1,2,\ldots,n)\equiv \frac{1}{\ab{1\,2}\ab{2\,3}\ab{3\,4}\cdots\ab{n\,1}}.\label{PT_factor_definition}}
Notice that each Parke-Taylor factor is cyclically symmetric; and so there are $(n\mi1)!$ distinct (but not independent) Parke-Taylor factors. In particular, for three particles there are two cyclically-distinct Parke-Taylor factors, $PT(1,2,3)$ and $PT(1,3,2)$, which differ from each other by an overall sign:\\[-6pt]
\eq{PT(1,2,3)=-PT(1,3,2).\label{sign_change_via_triple_reorientation}}

\newpage
\subsection{Geometry of Extended Positivity and Parke-Taylor Decompositions}\label{parke_taylor_sums_subsection}

Given the connection between on-shell forms and Grassmannian positivity for planar graphs, it is natural to ask whether the general non-planar on-shell forms also have some ``positive" interpretation. The obvious obstacle is that the notion of positivity seems to depend on ordering the columns of the matrix $C$, and there is no natural ordering beyond the planar limit. However as we will now see, associated with this, there are in fact many different ``positive regions" associated with a single on-shell diagram. Too see this, let us look at the black vertices of an MHV leading singularity---each of which contributes a factor equivalent to a $3$-particle Parke-Taylor factor involving the legs attached to it:
\vspace{-3pt}\eq{\hspace{-120pt}(\tau_1\,\tau_2\,\tau_3)\raisebox{-1pt}{\text{{\Large$\,\Rightarrow\!$}}} \int\!\!\frac{d\alpha_{\tau_1}\,d\alpha_{\tau_2}\,d\alpha_{\tau_3}}{\mathrm{vol}(GL(1))}\frac{\delta^{2\times1}\big(\lambda_{\tau_1}\alpha_{\tau_1}+\lambda_{\tau_2}\alpha_{\tau_2}+\lambda_{\tau_3}\alpha_{\tau_3}\big)}{\alpha_{\tau_1}\,\alpha_{\tau_2}\,\alpha_{\tau_3}}=\frac{1}{\ab{\tau_1\,\tau_2}\ab{\tau_2\,\tau_3}\ab{\tau_3\,\tau_1}}.\label{black_vertex_contribution_v2}\hspace{-100pt}\vspace{-0pt}}
Each black vertex can be seen to be in correspondence with the geometric constraint that $\{\lambda_{\tau_1},\lambda_{\tau_2},\lambda_{\tau_3}\}$ (viewed as point in $\mathbb{RP}^1$) are positive with respect to one of the two cyclic orderings of the three points. Thus, each triple $\tau\!\in\!T$ can be understood to impose the constraint that either $(\tau_1\,\tau_2\,\tau_3)$ or $(\tau_1\,\tau_3\,\tau_2)$ are cyclically ordered as points in $\mathbb{RP}^1$. To indicate how each black vertex is to be ordered, will henceforth distinguish between a triple denoted $(\tau_1\,\tau_2\,\tau_3)$ from one denoted $(\tau_1\,\tau_3\,\tau_2)$.

We can equivalently describe this picture as an extended positive region in the Grassmannian. Let us use ``little group rescaling" on each column of $C$, $c_a\!\mapsto\!t_a\,c_a$, to bring the columns to the projectivized form $c_a\!\mapsto\!\binom{1}{\widehat{c}_a}$. The positivity of a minor
$\det\{c_a,c_b\}\!>\!0$ simply means that $\widehat{c}_a\!<\!\widehat{c}_b$. Then, for every triple $(\tau_1,\tau_2,\tau_3)$ associated with our on-shell diagram, we can choose an ordering for the associated $(\widehat{c}_{\tau_1},\widehat{c}_{\tau_2},\widehat{c}_{\tau_3})$. Any (mutually consistent) choice of such orderings thus defines an extended positive region in the $G_{\mathbb{R}}(2,n)$, and our on-shell form has logarithmic singuarities on the boundaries of this region.

Although the choice of how each black vertex is cyclically ordered can change the final expression by at most an overall sign, it can significantly alter its geometric interpretation. For example, it is not hard to see that if every black vertex of a planar diagram were ordered consistently with the graph (all ordered clockwise, for example), then the cyclic ordering of all the $\lambda$'s will be completely fixed to match the ordering of legs around the boundary of the graph. In this case, the geometric region associated with the graph is simply the positive region of $G_{}(2,n)$ associated with the ordering of external legs around the graph and the resulting on-shell function must be a single Parke-Taylor factor. This can be seen in the following example:
\vspace{-6pt}\eq{\raisebox{-33pt}{\includegraphics[scale=.9]{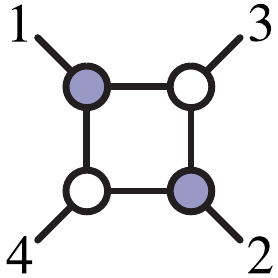}}\hspace{-5pt}\left\{\begin{array}{@{}c@{}}{\color{black}(1\,3\,4)}\\{\color{black}(3\,2\,4)}\end{array}\right\}\hspace{10pt}\raisebox{-2.85pt}{{\huge$\Leftrightarrow$}}\hspace{12.5pt}\begin{array}{@{}c@{}}\phantom{PT(1,2,3,4)}\\\left\{\raisebox{-25.25pt}{\includegraphics[scale=1]{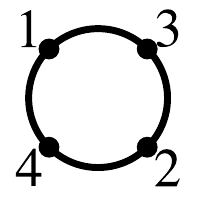}}\right\}\\PT(1,3,2,4)\end{array}\label{first_PT_example}\vspace{-2pt}}
Notice that the constraint that $(1\,3\,4)$ be cyclically ordered together with the constant that $(3\,2\,4)$ be cyclically ordered fixes the overall cyclic ordering of legs to be $(1\,3\,2\,4)$; and so all the poles of the leading singularity lie along the boundary of the positive region of $G(2,4)$ with the columns ordered according to $(1\,3\,2\,4)$.

However, the region of $G_{}(2,4)$ we associate with this on-shell function changes importantly if we choose to reorient either of the two black vertices. For example, by changing the orientation of the first black vertex from $(1\,3\,4)$ to $(1\,4\,3)$, the constraints would no longer fix the cyclic ordering of all the $\lambda$'s:
\vspace{-8pt}\eq{\raisebox{-33pt}{\includegraphics[scale=.9]{g24_graph_2}}\hspace{-5pt}\left\{\begin{array}{@{}c@{}}{\color{black}(1\,4\,3)}\\{\color{black}(3\,2\,4)}\end{array}\right\}\hspace{10pt}\raisebox{-2.85pt}{{\huge$\Leftrightarrow$}}\hspace{12.5pt}\begin{array}{@{}c@{}}\phantom{PT(1,3,4,2)}\\\hspace{0pt}\left\{\raisebox{-25.25pt}{\includegraphics[scale=1]{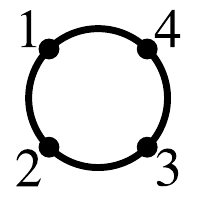}}\hspace{7.5pt}\mathrm{or}\hspace{7.5pt}\raisebox{-25.25pt}{\includegraphics[scale=1]{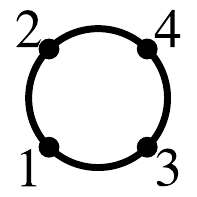}}\right\}\\PT(1,4,3,2)\hspace{2.5pt},\hspace{2.5pt}PT(1,2,4,3)\end{array}\vspace{-0pt}\label{eg_pt_sum_4_pt}}
As illustrated above, cyclically ordering the legs so that $(1\,4\,3)$ and $(3\,2\,4)$ are correctly ordered allows for {\it two} possible overall orderings: $(1\,4\,3\,2)$ or $(1\,2\,4\,3)$. Because the constraints on overall ordering are {\it weaker}, the region of $G(2,4)$ associated with this function is {\it larger} than a single positive region. And we now have a geometric basis for the identity:
\eq{-PT(1,3,2,4)=PT(1,4,3,2)+PT(1,2,4,3).}
(Here, the minus sign follows from the reorientation of a single triple according to, (\ref{sign_change_via_triple_reorientation}).)

More generally, there can be many orderings of the external legs consistent with a choice of orientation for each triple. When this is the case, the combined region will span more than one of the different positive parts of $G(2,n)$---those for which all the triples are cyclically ordered correctly. Because the boundary of the combined region exactly corresponds to the boundary determined by the triples, we know the volume form in (\ref{grassmannian_formula}) is cohomological to the positive sum of positroid volume forms. Expressed in terms of functions of the external kinematical data:
\vspace{5pt}\eq{\framebox[0.4\textwidth]{$\displaystyle\hspace{-30pt} \widetilde{f}_\Gamma=\sum_{\fwboxL{0pt}{\hspace{-10pt}\{\sigma\!\in\!(\mathfrak{S}_n/\mathbb{Z}_n)|\forall \tau\!\in\!T\!\!:\!\sigma_{\tau_1}\!\!<\!\sigma_{\tau_2}\!\!<\!\sigma_{\tau_3}\}}}\hspace{-00pt}PT(\sigma_1,\ldots,\sigma_n),$}\label{parke_taylor_decomposition_formula}}
where $\mathfrak{S}_n/\mathbb{Z}_n$ denotes the set of cyclically-inequivalent permutations, and the sum is taken over all $\sigma\!\in\!(\mathfrak{S}_n/\mathbb{Z}_n)$ for which each triple $\tau\!\in\!T$ is cyclically ordered correctly.

This identity is quite non-trivial. Let's illustrate it for the $6$-particle leading singularity written explicitly in (\ref{g26_example_formula}). For one choice of how to order each triple, we see that the compact expression becomes the sum of seven Parke-Taylor factors as follows:
\vspace{-17pt}{\small \eq{\hspace{-250pt}\begin{array}{@{}c@{}}\phantom{T\equiv\{\}}\\\fwboxL{135pt}{\raisebox{-65pt}{\includegraphics[scale=0.9]{g26_graph}}}\\\hspace{-50pt}T\!\equiv\!\{(1\,2\,3),(2\,5\,6),(3\,4\,6),(4\,5\,1)\}\phantom{\!\equiv\!T}\hspace{-50pt}\end{array}\begin{array}{@{}l@{}l@{}l@{}}~\\~\\~\\=&\phantom{+}&PT(1,2,3,4,5,6)+PT(1,2,4,5,6,3)\\&+&PT(1,4,2,5,6,3)+PT(1,4,5,6,2,3)\\&+&PT(1,4,6,2,3,5)+PT(1,4,6,2,5,3)\\&+&PT(1,6,2,3,4,5)\end{array}\hspace{-200pt}\label{pt_expansion_for_6_pt_example}}}
\hspace{-4pt}Notice that each of the triples $\tau\!\in\!T$ are cyclically ordered correctly in each of the Parke-Taylor factors above. Changing the orientation of the last two triples, we could have labeled the diagram by $T'\!\equiv\!\{(1\,2\,3),(2\,5\,6),(3\,6\,4),(4\,1\,5)\}$, resulting in the Parke-Taylor decomposition:
\eq{\begin{array}{@{}l@{}l@{}l@{}l@{}}&PT(1,2,3,5,6,4)+PT(1,2,5,3,6,4)+PT(1,2,5,4,3,6)+PT(1,2,5,6,4,3)\\+&PT(1,5,6,2,4,3)+PT(1,5,6,4,2,3)+PT(1,6,2,5,4,3).\end{array}}

Notice that each of the choices for how to orient the triples produces a different expression of the form (\ref{parke_taylor_decomposition_formula}). Moreover, each of these formulae connects the leading singularity with a different region within the Grassmannian---providing a geometric basis for a large number of relations satisfied by Parke-Taylor amplitudes involving different leg orderings.

\subsection{Geometry of $U(1)$ Decoupling and KK Relations}\label{kk_and_u1_decoupling_subsection}
As we have seen in the example (\ref{eg_pt_sum_4_pt}), even a planar diagram can be decomposed into a sum of different Parke-Taylor factors if the black vertices are oriented in a way contrary to the planar ordering; and the equivalence of these different formulae for the same on-shell function follow geometrically from the fact that they correspond to cohomologically equivalent volume forms on the Grassmannian. Indeed, it turns out that the $U(1)$ decoupling identities and KK relations \cite{Kleiss:1988ne} can both be given a geometric interpretation in this way. We can also understand these relations as residue theorems which can be graphically represented as a sum over certain deleted edges of the on-shell diagram, as has been observed independently by Andrew Hodges, \cite{AndrewHodges}.

Although the $U(1)$ decoupling identities are of course included among the KK relations, they are often presented in different ways. And so for the sake of clarity and illustration, it is useful to consider the two cases separately. In their most familiar form, the $U(1)$ decoupling identities can be seen to follow from the equivalence of two representations of the following planar diagram (with different orientations chosen for one black vertex):
\vspace{-5pt}\eq{\hspace{-190pt}\raisebox{-68.5pt}{\includegraphics[scale=.9]{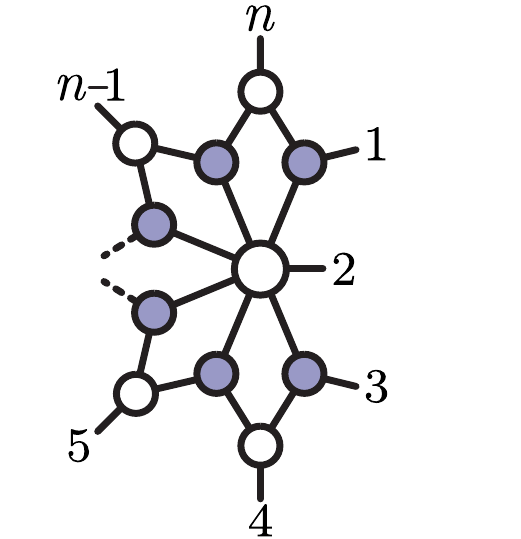}}\hspace{-20pt}\begin{array}{@{}l@{$\hspace{-8pt}$}}\phantom{\Big(PT(12\cdots n)}\\[6pt]%
\fwbox{0pt}{(-1)\hspace{15pt}}\left\{\!\begin{array}{@{$($}c@{$\,$}c@{$\,$}c@{$)$}}2&n&1\\2&3&4\\2&4&5\\[-5pt]\multicolumn{3}{c}{\vdots}\\[-2pt]2&n\!\mi\!1&n\end{array}\!\right\}\\[10pt]\hline\\[-12pt] \fwboxL{0pt}{\!\!\!\!\!\hspace{-10pt}-PT(1,2,\ldots,n)\phantom{}}\end{array}\hspace{5.5pt}\raisebox{-2pt}{$=\;$}\begin{array}{@{}l@{$\hspace{-9pt}$}}\phantom{\Big(PT(12\cdots n)}\\[6pt]%
\left\{\!{\color{deemph}\begin{array}{@{$($}c@{$\,$}c@{$\,$}c@{$)$}}\multicolumn{1}{@{${\color{black}(}$}c@{$\,$}}{{\color{black}2}}&{\color{hred}1}&\multicolumn{1}{@{}c@{${\color{black})}$}}{{\color{hred}n}}\\2&3&4\\2&4&5\\[-5pt]\multicolumn{3}{c}{\vdots}\\[-2pt]2&n\!\mi\!1&n\end{array}}\!\right\}\\[10pt]\hline\\[-12pt] \hspace{-11.5pt}=\;\fwboxL{0pt}{\!PT(1,n,2,\ldots,n\!\mi\!1)\pl\!\ldots\!\pl\,\!PT(1,3,\ldots,n,2).}\end{array}\vspace{-2pt}\label{u1_decoupling_identity}}

Similarly, we can derive the KK relations in their standard form by choosing different orientations for the black vertices of the following planar diagram:
\vspace{-15pt}\eq{\raisebox{-68.5pt}{\includegraphics[scale=1]{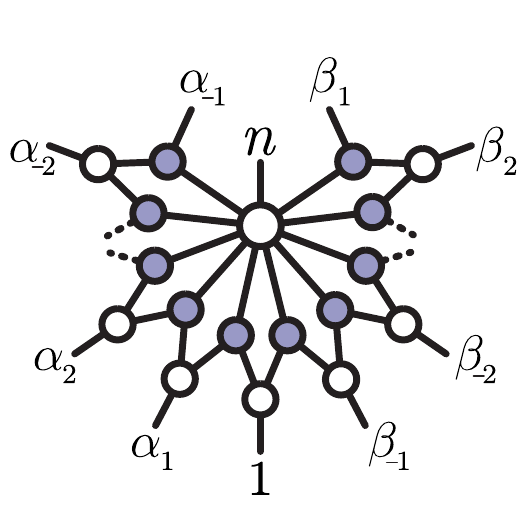}}\hspace{5pt}\left\{\!\begin{array}{@{$($}c@{$\,$}c@{$\,$}c@{$)$}}1&\alpha_{\phantom{\text{-}}1}&n\,\\\alpha_{\phantom{\text{-}}1}&\alpha_{\phantom{\text{-}}2}&n\,\\[-5pt]\multicolumn{3}{c}{\vdots}\\[-2pt]\alpha_{\text{-}2}&\alpha_{\text{-}1}&n\\n&\beta_{\phantom{\text{-}}1}&\beta_{\phantom{\text{-}}2}\\[-5pt]\multicolumn{3}{c}{\vdots}\\[-2pt]n&\beta_{\text{-}2}&\beta_{\text{-}1}\\n&\beta_{\text{-}1}&1\end{array}\!\right\}=(-1)^{n_\beta}\left\{\!{\color{deemph}\begin{array}{@{$($}c@{$\,$}c@{$\,$}c@{$)$}}1&\alpha_{\phantom{\text{-}}1}&n\,\\\alpha_{\phantom{\text{-}}1}&\alpha_{\phantom{\text{-}}2}&n\,\\[-5pt]\multicolumn{3}{c}{\vdots}\\[-2pt]\alpha_{\text{-}2}&\alpha_{\text{-}1}&n\\\multicolumn{1}{@{${\color{black}(}$}c@{$\,$}}{{\color{black}n}}&{\color{hred}\beta_{\phantom{\text{-}}2}}&\multicolumn{1}{@{}c@{${\color{black})}$}}{{\color{hred}\beta_{\phantom{\text{-}}1}}}\\[-5pt]\multicolumn{3}{c}{\vdots}\\[-2pt]\multicolumn{1}{@{${\color{black}(}$}c@{$\,$}}{{\color{black}n}}&{\color{hred}\beta_{\text{-}1}}&\multicolumn{1}{@{}c@{${\color{black})}$}}{{\color{hred}\beta_{\text{-}2}}}\\\multicolumn{1}{@{${\color{black}(}$}c@{$\,$}}{{\color{black}n}}&{\color{hred}1}&\multicolumn{1}{@{}c@{${\color{black})}$}}{{\color{hred}\beta_{\text{-}1}}}\end{array}\!}\right\}.\vspace{-2pt}}
Notice that this relates a planar diagram involving legs ordered $(1,\alpha_1,\ldots,\alpha_{\text{-}1},n,\beta_1,\ldots,\beta_{\text{-}1})$ a sum of Parke-Taylor factors:
\eq{(-1)^{n_\beta}\times PT(1,\alpha_1,\ldots,\alpha_{\text{-}1},n,\beta_{1},\ldots,\beta_{\text{-}1})=\!\!\!\sum_{\fwboxL{30pt}{\,\,\,\,\,\,\sigma\!\in\!\big(\{\alpha_1,\ldots,\alpha_{\text{-}1}\}\!\shuffle\!\{\beta_{\text{-}1},\ldots,\beta_1\}\big)}}PT(1,\sigma_1,\ldots,\sigma_{n\text{-}2},n).\label{kk_relations}}

\subsection{Two Views of MHV Leading Singularities}

We have now described two quite different ways to represent MHV leading singularities in terms of combinatorial data describing the on-shell diagram: the compact formula given in equation (\ref{freddys_formula_for_general_diagrams}), and as the positive sum of Parke-Taylor factors given in equation (\ref{parke_taylor_decomposition_formula}). Each of these representations makes different aspects of the functions manifest. In particular, the compact expression (\ref{freddys_formula_for_general_diagrams}), makes all of the physical singularities manifest at the cost of obscuring the fact that all these singularities are strictly logarithmic; in contrast, the Parke-Taylor expansion (\ref{parke_taylor_decomposition_formula}) makes manifest the fact that all singularities are logarithmic, but at the cost of introducing spurious boundaries which join each of the positive regions.

This distinction can be illustrated in the case of the $6$-particle leading singularity given above in (\ref{g26_example_formula}). Taking the residue about $\ab{12}\!\to\!0$ by sending $\lambda_1\!\rightsquigarrow\!\lambda_2$ results in the expression:
\vspace{-0pt}\eq{\hspace{-100pt}\frac{\big(\!\ab{34}\ab{51}\ab{62}\pl\ab{14}\ab{25}\ab{63}\!\big)^2}{\ab{12}\hspace{-1pt}\ab{23}\hspace{-1pt}\ab{31}\hspace{-1pt}\ab{25}\hspace{-1pt}\ab{56}\hspace{-1pt}\ab{62}\hspace{-1pt}\ab{34}\hspace{-1pt}\ab{46}\hspace{-1pt}\ab{63}\hspace{-1pt}\ab{45}\hspace{-1pt}\ab{51}\hspace{-1pt}\ab{14}}\!\!\xrightarrow[\!\!\lambda_1\rightsquigarrow\lambda_2\!\!]{}\!\!\frac{\big(\!\ab{34}\ab{52}\ab{62}\pl\ab{24}\ab{25}\ab{63}\!\big)^2}{\ab{23}^2\hspace{-1pt}\ab{25}^2\hspace{-1pt}\ab{56}\hspace{-1pt}\ab{62}\hspace{-1pt}\ab{34}\hspace{-1pt}\ab{46}\hspace{-1pt}\ab{63}\hspace{-1pt}\ab{45}\hspace{-1pt}\ab{24}}.\nonumber\label{example_dangerous_residue}\hspace{-100pt}}
The presence of the double poles $\ab{23}^2$ and $\ab{2\,5}^2$ in the denominator above would seem to suggest non-logarithmic behavior. However, closer inspection indicates that the numerator $\Psi$ factorizes in the limit $\lambda_1\!\rightsquigarrow\!\lambda_2$. Making use of Schouten relations, the numerator becomes:\\[-8pt]
\eq{\big(\!\ab{34}\ab{52}\ab{62}\pl\ab{24}\ab{25}\ab{63}\!\big)^2=\big(\ab{23}\ab{25}\ab{46}\big)^2.}
The ability of the numerator $\Psi$ to correctly cancel all apparent double poles is quite remarkable, but follows from the fact that its representation in the Grassmannian, (\ref{grassmannian_formula}), involves a measure with only logarithmic singularities.

Of course, the fact that all the iterated singularities of (\ref{example_dangerous_residue}) are purely logarithmic is made clear by its representation in terms of Parke-Taylor factors, (\ref{pt_expansion_for_6_pt_example}). However, the Parke-Taylor expansion introduces spurious boundaries which are cancelled only in combination. For example, both $PT(1,2,3,4,5,6)$ and $PT(1,6,2,3,4,5)$ have a pole when $\ab{16}\!\to\!0$; but this does not correspond to any singularity of the on-shell function, and the corresponding residue cancels between the two terms.

\section{Concluding Remarks}\label{conclusions_section}
We have seen that non-planar MHV leading singularities can always be written as a sum of the planar ones with different orderings of legs. Already for the case of $k\!=\!3$ and $n\!=\!6$, we have checked that this is no longer the case---not all leading singularities can be decomposed into differently ordered planar ones, and new objects appear. But since we know on general grounds that there are a finite number of objects that can appear for any $(n,k)$, it is likely that a complete combinatorial characterization of them should be possible. It is for instance clear that by cutting enough legs, any non-planar on-shell diagram can be made planar (as in \cite{Chen:2014ara}), and this is one avenue towards a non-planar classification.

We have also given a geometric interpretation of the $U(1)$ decoupling and KK relations for Parke-Taylor factors. For general $k$, these relations (as well the BCJ relations) are usually associated with statements about complete amplitudes; it would be interesting to explore whether they have on-shell avatars for general $k$, with an interpretation in terms of Grassmannian geometry.

Finally, it is natural to combine the results of this note with the conjecture of \mbox{ref.\ \cite{Arkani-Hamed:2014via}}, that non-planar loop amplitude integrands have logarithmic singularities and can be expressed in ``$d\!\log$-form''. Since all the leading singularities can be written as a linear combination of Parke-Taylor factors, we expect that the integrand can be represented as $d\!\log$-forms in the loop momentum variables, multiplied by color factors and Parke-Taylor amplitudes. This suggests that after integration, the non-planar MHV amplitudes in $\mathcal{N}\!=\!4$ SYM at all loop orders are expressible as a sum of polylogarithms weighted by color and (unordered) Parke-Taylor factors.

\acknowledgments We are grateful to Andrew Hodges and Lauren Williams for helpful discussions. N.~A.-H. is supported by the Department of Energy under grant number DE-FG02-91ER40654; J.~L.~B. is supported by a MOBILEX grant from the Danish Council for Independent Research; F.~C. is supported by the Perimeter Institute for Theoretical Physics which is supported by the Government of Canada through Industry Canada and by the Province of Ontario through the Ministry of Research \& Innovation; and J.~T. is supported in part by the David and Ellen Lee Postdoctoral Scholarship and by the Department of Energy under grant number DE-SC0011632.

\providecommand{\href}[2]{#2}\begingroup\raggedright\endgroup

\end{document}